\documentclass{article}
\usepackage[final,nonatbib]{neurips_2020}
\usepackage{url}
\usepackage{times}
\usepackage{graphicx}
\usepackage{amsmath,amssymb}
\usepackage{amsthm}
\usepackage{algorithm}
\usepackage{algpseudocode}
\usepackage{epsfig}
\usepackage{fullpage}
\usepackage{bbm}
\usepackage{subcaption}

\usepackage{float}

\newtheorem{remark}{Remark}

\title{Matching through Embedding in Dense Graphs}
\author{%
  Nitish K. Panigrahy \\
  University of Massachusetts Amherst\\
  Amherst, MA 01003 \\
  \texttt{nitish@cs.umass.edu} \\ 
  \And 
  Prithwish Basu\\
  Raytheon BBN Technologies\\
  Cambridge, MA 02138\\
  \texttt{prithwish.basu@raytheon.com}\\
  \And
  Don Towsley\\
  University of Massachusetts Amherst\\
  Amherst, MA 01003 \\
  \texttt{towsley@cs.umass.edu} \\
} 
\begin{document}

\maketitle

\begin{abstract}
Finding optimal matchings in dense graphs is of general interest and of particular importance in social,  transportation and biological networks. While developing optimal solutions for various matching problems is important, the running times of the fastest available optimal matching algorithms are too costly.  However, when the vertices of the graphs are point-sets in $\pmb{\mathbb{R}^d}$ and edge weights correspond to the euclidean distances, the available optimal matching algorithms are substantially faster. In this paper, we propose a novel \emph{network embedding} based heuristic algorithm to solve various matching problems in dense graphs. In particular, using existing network embedding techniques, we first find a low dimensional representation of the graph vertices in $\pmb{\mathbb{R}^d}$ and then run faster available matching algorithms on the embedded vertices. To the best of our knowledge, this is the first work that applies network embedding to solve various matching problems. Experimental results validate the efficacy of our proposed algorithm.
\end{abstract}

\section{Introduction}\label{sec:intro}
Matching in graphs is one of the most well-studied problems in economics and computer science. This is an ubiquitous problem with application areas ranging from transportation networks \cite{Aggarwal95} to social networks  \cite{Chu2013}, and  computational chemistry \cite{Frohlich05}. For instance, in the case of social networks, tag suggestions and localization problems for images can be transformed into a weighted bipartite graph matching problems \cite{Chu2013}. Similarly, the processes of de-anonymization and privacy inference in social networks  can be reduced to finding the maximum weighted bipartite matching of the corresponding knowledge graph \cite{Qian2016}.


Given a graph $G = (V,E)$, a matching $M$ is a subset of edges such that no two edges in $M$ share a common vertex. The matching $M$ is \emph{perfect} if every $v \in V$ belongs to an edge of $M$. For weighted graphs, it is often required to compute a perfect matching that is optimal with respect to some criterion. Given a real weight $w_e$ for each edge $e \in E,$ a \emph{minimum cost} matching (MCM) minimizes $\sum_{e \in M} w_e$ among all feasible perfect matchings $M$ for $G.$ Similarly, a bottleneck matching (BM) minimizes $\max_{e \in M} w_e$ while a \emph{uniform matching} (UM) minimizes $\max_{e \in M} w_e - \min_{e \in M} w_e.$ A \emph{minimum deviation} matching (MDM) minimizes $(1/|V|\sum_{e \in M} w_e) - \min_{e \in M} w_e.$


A plethora of work has been done to develop efficient algorithms for obtaining optimal solutions for various matchings. However, for many real-world problems with larger graphs \cite{Beier00, Monien00}, the running times of the fastest available matching algorithms are too costly. For example, the best known algorithm for the minimum cost matching problem runs in $O(n^3)$ time for a dense graph with  $n$ vertices \cite{Kuhn55}. For massive graphs, this is quite inefficient. Thus designing efficient approximate algorithms permit the solution of large instances of matching problems that arise in practical situations.

\begin{table*}[htbp]
\hspace{-1.5cm}
\begin{tabular}{|c| c| c|c|c|}
\hline
{\bf Graph Type}&{\bf Matching Type}&{\bf Algo.} &  {\bf Algo. Type}& {\bf Complexity}\\
\hline
Bipartite&Minimum Cost (MCM)&Kuhn et al. \cite{Kuhn55}&Non-Euclidean&$O(n^{3})$\\
&&Agarwal et al. \cite{Agarwal95}&\bf{Euclidean}-$\pmb{\mathbb{R}^d}$&$\pmb{O(n^{2+\epsilon})}$\\
\hline
Bipartite&Bottleneck (BM)&Punnen et al. \cite{Punnen94}&Non-Euclidean&$O(n^2\sqrt{n})$\\
&&Efrat et al. \cite{Efrat01}&\bf{Euclidean}-$\pmb{\mathbb{R}^d}$&$\pmb{O(n^{1.5}\log^dn)}$\\
\hline
Bipartite&Uniform (UM)&Martello et al. \cite{Martello84}&Non-Euclidean&$O(n^{4})$\\
&&Efrat et al. \cite{Efrat01}&\bf{Euclidean}&$\pmb{O(n^{10/3})}$\\
\hline
Bipartite&Minimum Deviation (MDM)&Efrat et al. \cite{Efrat98}&Non-Euclidean&$O(n^{4})$\\
&&Efrat et al. \cite{Efrat01}&\bf{Euclidean}&$\pmb{O(n^{10/3})}$\\
\hline
Non-bipartite&Minimum Cost (MCM) &Gabow et al. \cite{Gabow90}&Non-Euclidean&$O(nm + n^2\log n)$\\
&&Varadarajan et al. \cite{Varadarajan98}&\bf{Euclidean}-$\pmb{\mathbb{R}^d}$&$\pmb{O(n^{1.5}\log n)}$\\
\hline
Non-bipartite&Bottleneck (BM)&Gabow et al. \cite{Gabow88}&Non-Euclidean&$O(m\sqrt{n\log n})$\\
&&Efrat et al. \cite{Efrat00}&\bf{Euclidean}-$\pmb{\mathbb{R}^d}$&$\pmb{O\left(n^{2 - 2/(\lceil d/2\rceil+1)+\epsilon}\right)}$\\
\hline
\end{tabular}
\caption{Running time complexities of various optimal matching algorithms.}
\label{tbltraff}
\end{table*}
There exist many constant factor approximation algorithms for the maximum weight matching (MWM) problem. However, there is no such algorithm for the minimum cost matching (MCM) problem. The greedy heuristic for the MCM problem, attempts to construct a minimum cost perfect matching by starting with an empty matching and iteratively adding a minimum weight edge between two exposed nodes. While the running time complexities of various greedy matching algorithms are generally linear in terms of the number of edges ($m$), the corresponding approximation ratios are high. The greedy heuristic runs in $O(m\log n)$ time and finds a solution with cost at most $4/3 \;n^{\log 3/2}$ times the optimum cost \cite{Reingold81}. \cite{Grigoriadis88} developed an $O(m)$ heuristic that constructs a matching with cost at most $2n^{\log_3 7/3}$ times the optimum cost.  Thus one should focus on designing efficient approximate algorithms with good performance guarantees.



While for non-Euclidean graphs the running time complexities of optimal matching algorithms are high, the available optimal matching algorithms are substantially faster for the Euclidean case, i.e. when the vertices of the graph are point sets in $\mathbb{R}^d$ and edge weights corresponds to euclidean distances. For example, the best known algorithm for the bottleneck matching problem runs in $O(n^{1.5}\log^d n)$ time for a bipartite Euclidean graph as compared to an $O(n^{2.5})$ for its non-Euclidean counterpart. Thus the following natural question arises. \emph{Can we leverage Euclidean matching techniques to obtain near-optimal solutions to non-Euclidean matching problems?}

In this work, we propose a \emph{network embedding} based algorithm to obtain approximate solutions to non-Euclidean matching problems. More precisely, using existing linear time network embedding techniques, we embed the vertices of the non-Euclidean graph into points in $\mathbb{R}^d$ such that the neighborhood of the vertices are approximately preserved. We then run faster available Euclidean matching algorithms on the embedded vertices. To the best of our knowledge, this is the first work that applies network embedding to solve various matching problems. Empirical results show the efficacy of our proposed algorithm.

\section{Technical Preliminaries}\label{sec:gm}
We consider the problem of finding approximate solutions to optimal matching problem in complete bipartite and general graphs\footnote{We assume the graph to be sufficiently dense for the case when it is not complete. In such a case, we can add $O(1)$ edges each with infinite costs to the graph to make it complete without affecting the overall running time complexity.}. We denote the graph as $G = (V,E)$ with $E \subset V \times V.$ Here, $V$ denotes the vertex set and $E$ denotes the edge set. We assign a real weight $w_e$ to each edge $e \in E.$ A matching $M \subset E$ is defined to be a set of edges such that no vertex of $G$ is incident to more than one edge of $M$. Denote $|V| = n$ and $|E| = m.$ Our goal is to compute a perfect matching that is near-optimal with respect to optimality criterion defined in the previous Section. 
%
\subsection{Geometry Helps in Matching}
A summary of the running time complexities of various optimal matching algorithms in Euclidean and non-Euclidean setting is given in Table \ref{tbltraff}. Note that, bipartite matching is a special case of general graph matching. Computing an optimal bipartite matching is more challenging than computing an optimal matching on a complete non-bipartite graph in a Euclidean setting. For example, MCM on a set of $2n$ points can be computed in $O(n^{1.5}\log n)$ time, while the best known algorithm for computing MCM between two point sets of size $n$ each takes  $O(n^{2+\epsilon})$ time. Also note that, there is considerable amount of savings in terms of running times in the Euclidean setting as compared to the non-Euclidean setting. For instance, MCM on a non-Euclidean non-bipartite complete graph runs in $O(n^3)$ time where as MCM on a Euclidean non-bipartite complete graph runs in $O(n^{1.5}\log n)$ time. Similar savings can be observed for many other matching problems, such as BM, UM and MDM. Below we transform the non-Euclidean matching problem into a Euclidean one through network embedding.



\section{Our Approach: Matching through Embedding}\label{sec:emb}
Given a graph $G$, we aim to learn a \emph{representation} of $V$ in a low-dimensional vector space $\mathbb{R}^d$, i.e., find a map $f: V \to \mathbb{R}^d$ such that the neighborhoods of nodes are approximately preserved. This allows us to execute matching algorithms on the set of vectors instead of doing that on $G$ itself. Below we present two network embedding techniques available from literature.
\begin{itemize}
\item{\it Deep Walk} \cite{Perozzi2014}: As a homogeneous network embedding method, DeepWalk performs uniform random walks to get a corpus of vertex sequences. Then the word2vec is applied on the corpus to learn vertex embeddings.
\item{\it Node2Vec} \cite{Grover2016}: This method extends DeepWalk by performing biased random walks to generate the corpus of vertex sequences. The hyper-parameters $p$ and $q$ can be set to different values.
\end{itemize}

We construct the embedded graph $G^\prime(V^\prime, E^\prime)$ as follows. We let $V^\prime = \{f(v), \forall v \in V\}.$ Also, we let $E^\prime = \{(f(u), f(v)), \forall (u,v) \in E\}.$ We then define weight of an edge $e^\prime = (u^\prime,v^\prime) \in E^\prime$ as $w_{e^\prime} = ||u^\prime - v^\prime||_2.$ Clearly $G^\prime$ is a Euclidean graph in dimension $d.$ We then apply available optimal Euclidean matching algorithms on $G^\prime$ to obtain a matching $M^\prime.$ We output $M^\prime$ as the approximate solution to the optimal matching problem on $G.$

\begin{remark}
Note that, the above mentioned embedding techniques can learn the representation of $V$ in $O(n\log n)$ time\cite{Chen2018}. Thus the overall time complexity of the proposed algorithm is equal to that of the corresponding Euclidean matching algorithm. 
\end{remark}

\section{Empirical Results}\label{sec:emp}
In this section we present experiments on several synthetic datasets. 
 \begin{figure}
\centering
\includegraphics[width=10cm, height=1cm]{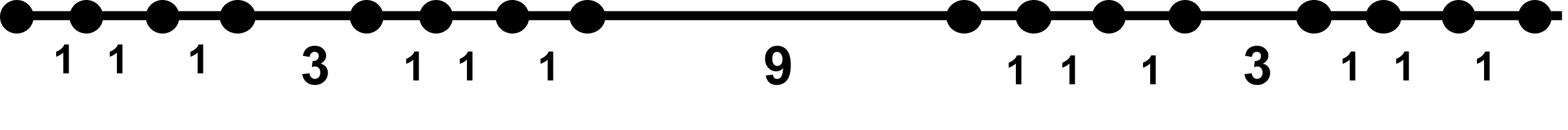}
\caption{Adversarial model for assigning edge weights with $n = 16$ \cite{Reingold81}.}
\label{greedy-adversarial}
\vspace{-0.1in}
\end{figure}
\begin{figure*}[htbp!]
\centering
\begin{minipage}{.32\textwidth}
\centering
\includegraphics[width=1\linewidth]{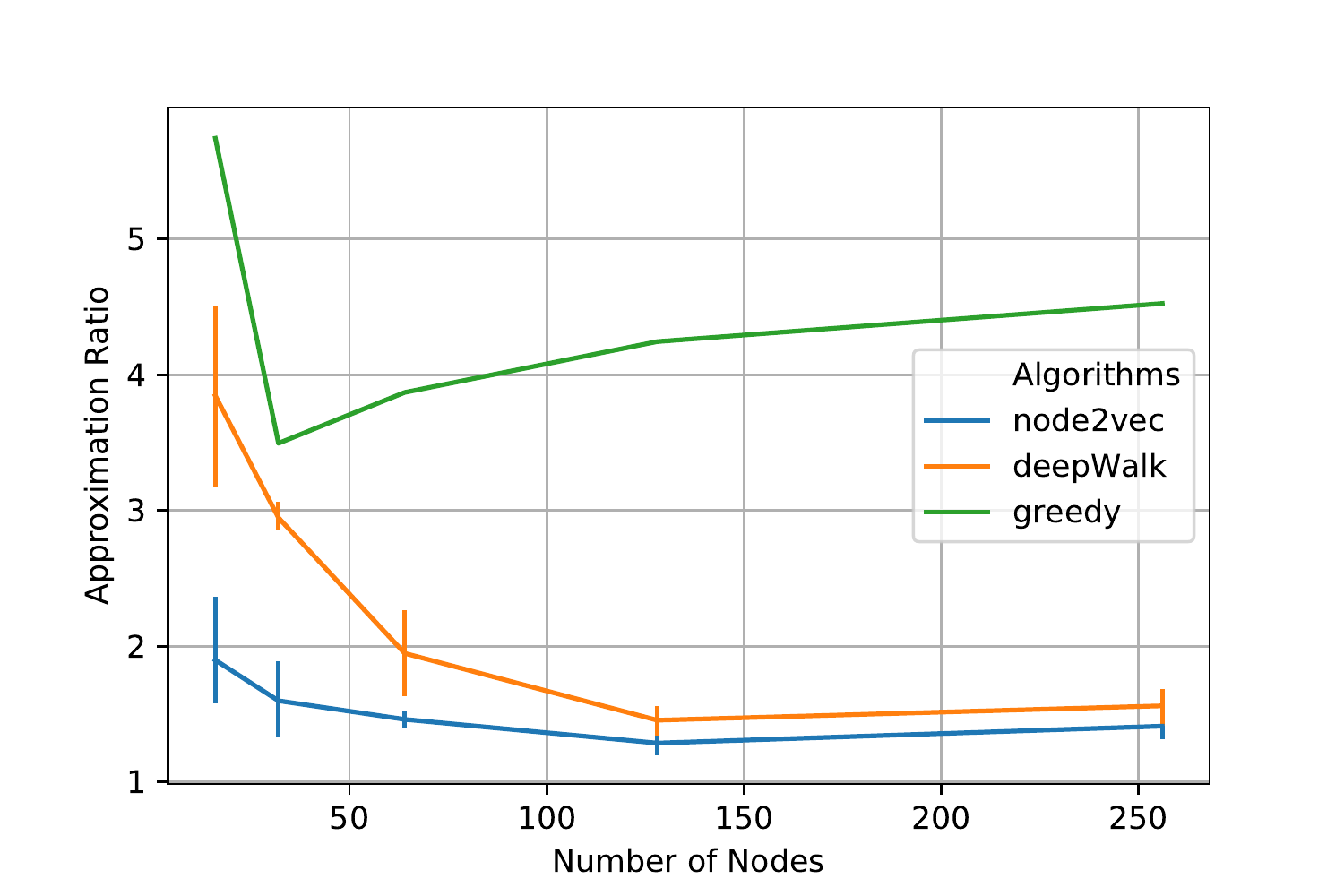}
\subcaption{}
\end{minipage}\hfill
\begin{minipage}{.32\textwidth}
\centering
\includegraphics[width=1\linewidth]{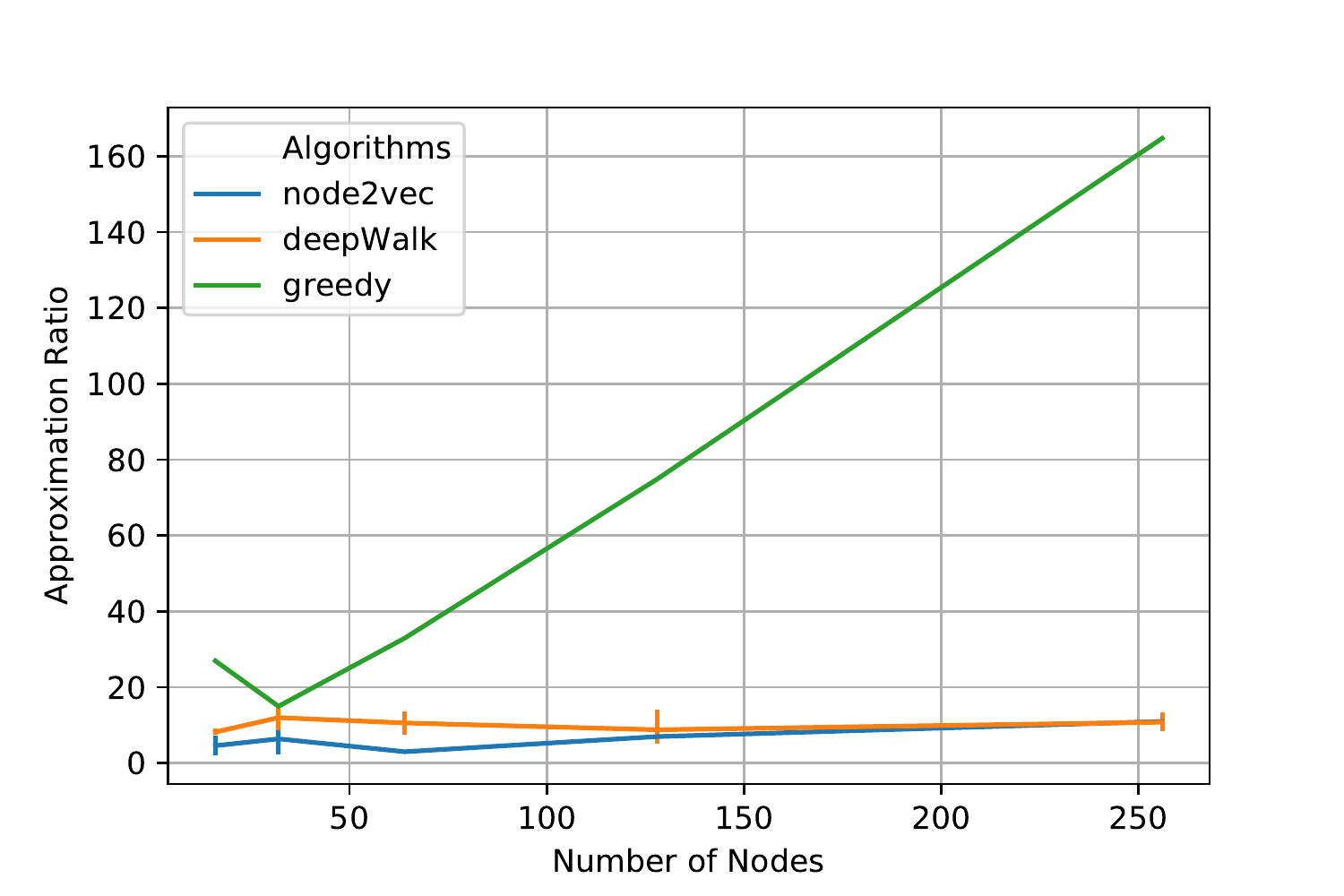}
\subcaption{}
\end{minipage}\hfill
\begin{minipage}{.32\textwidth}
\centering
\includegraphics[width=0.8\linewidth]{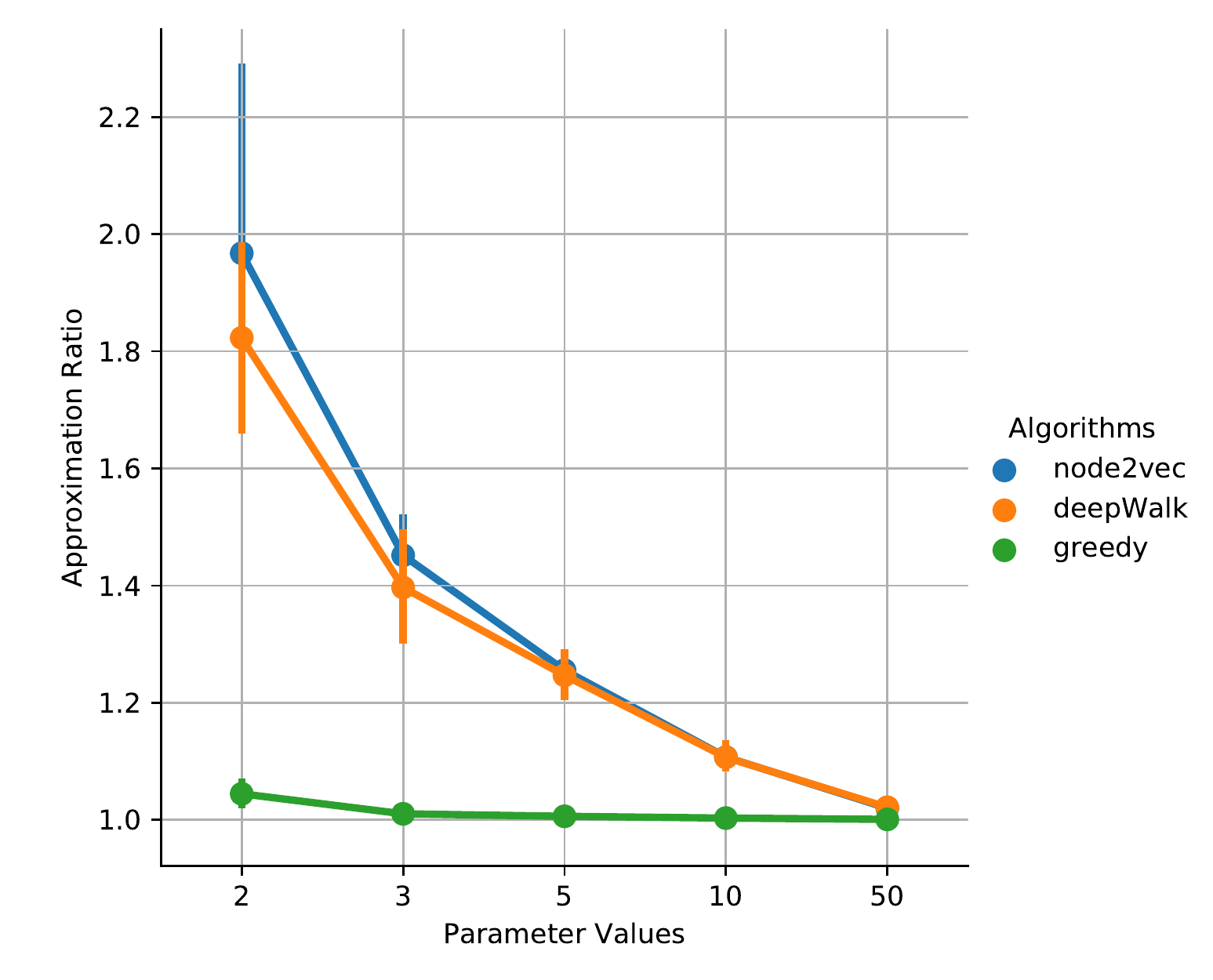}
\subcaption{}
\end{minipage}
\vspace{-0.5cm}
\begin{minipage}{.32\textwidth}
\centering
\includegraphics[width=1\linewidth]{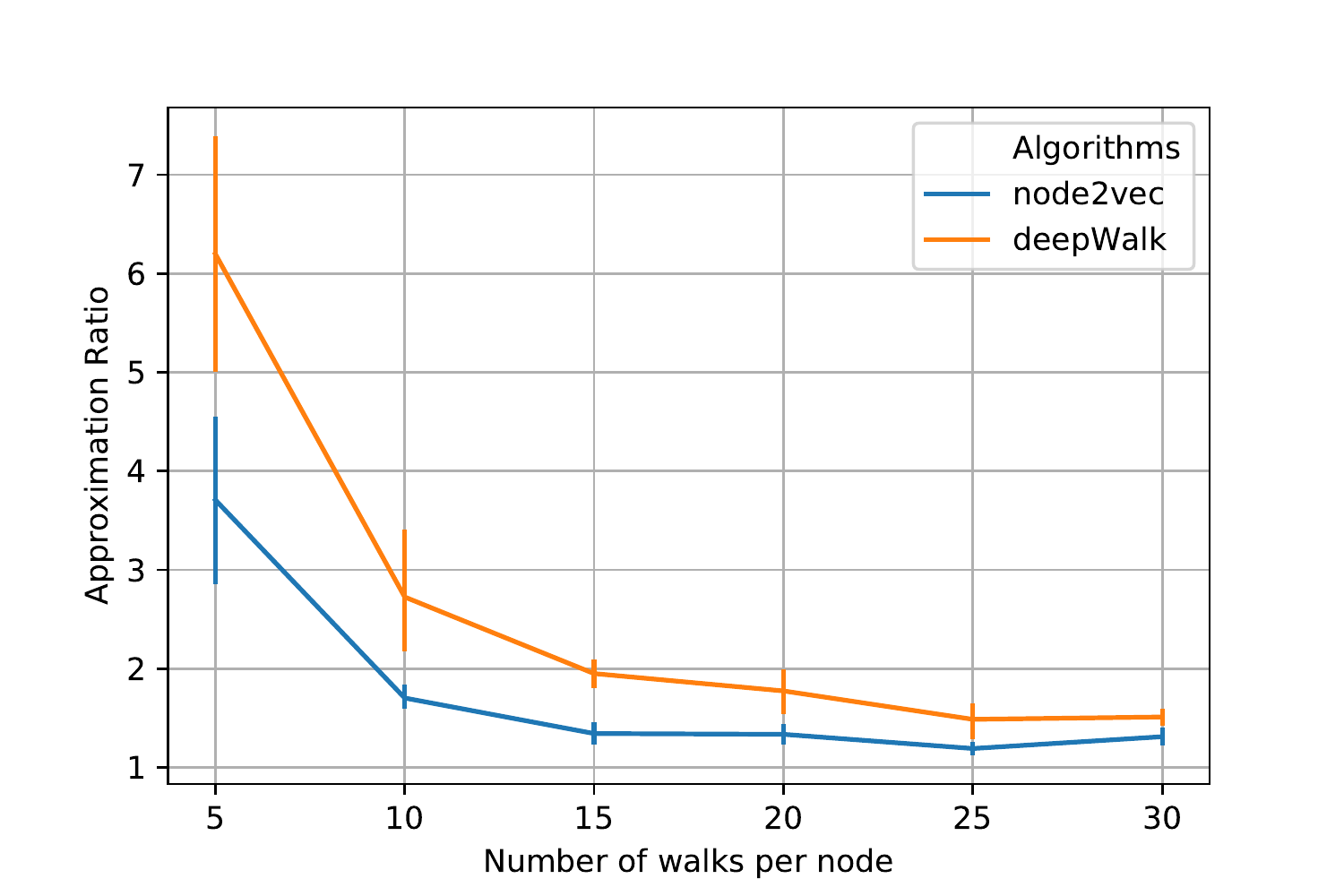}
\subcaption{}
\end{minipage}\hfill
\begin{minipage}{.32\textwidth}
\centering
\includegraphics[width=1\linewidth]{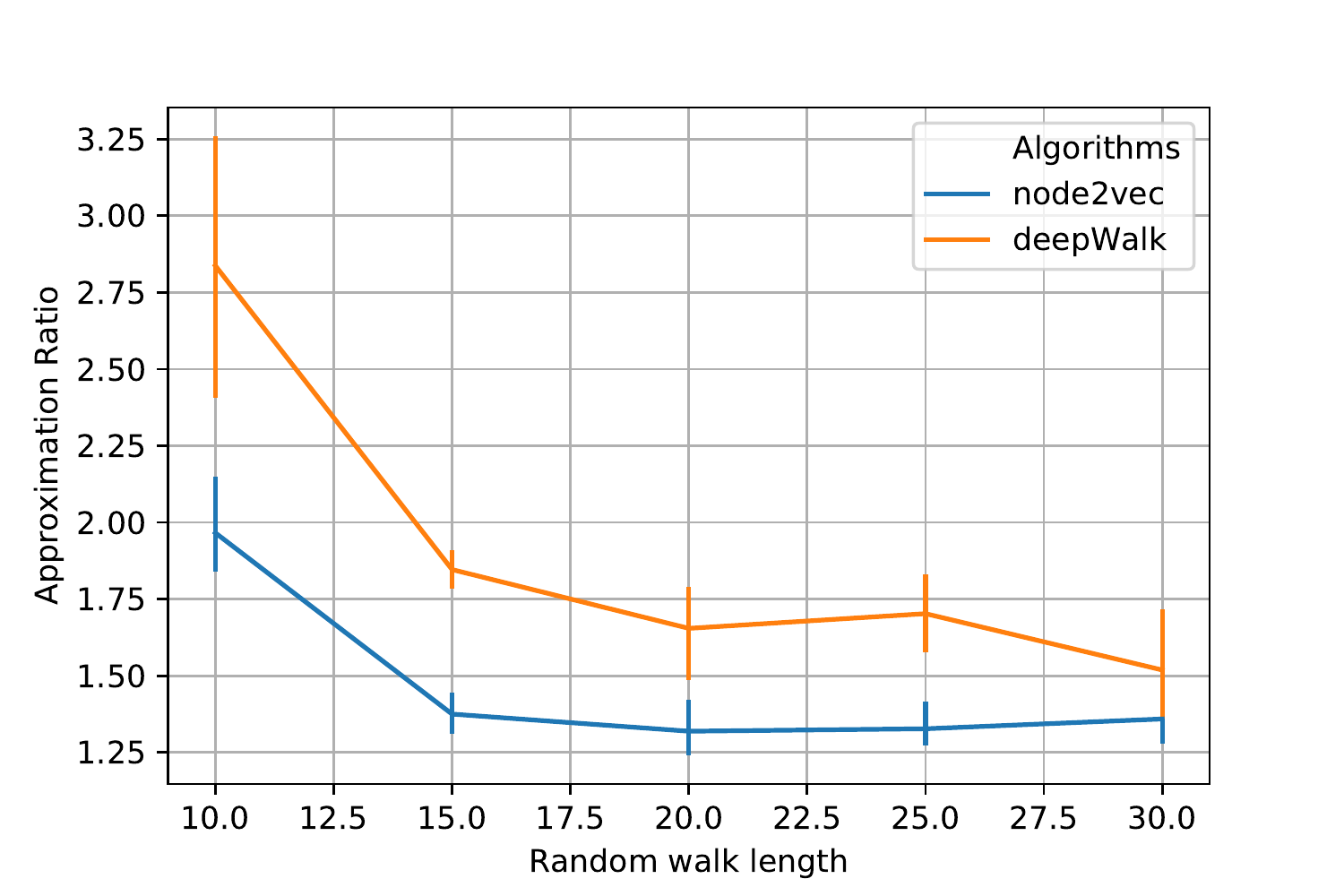}
\subcaption{}
\end{minipage}\hfill
\begin{minipage}{.32\textwidth}
\centering
\includegraphics[width=1\linewidth]{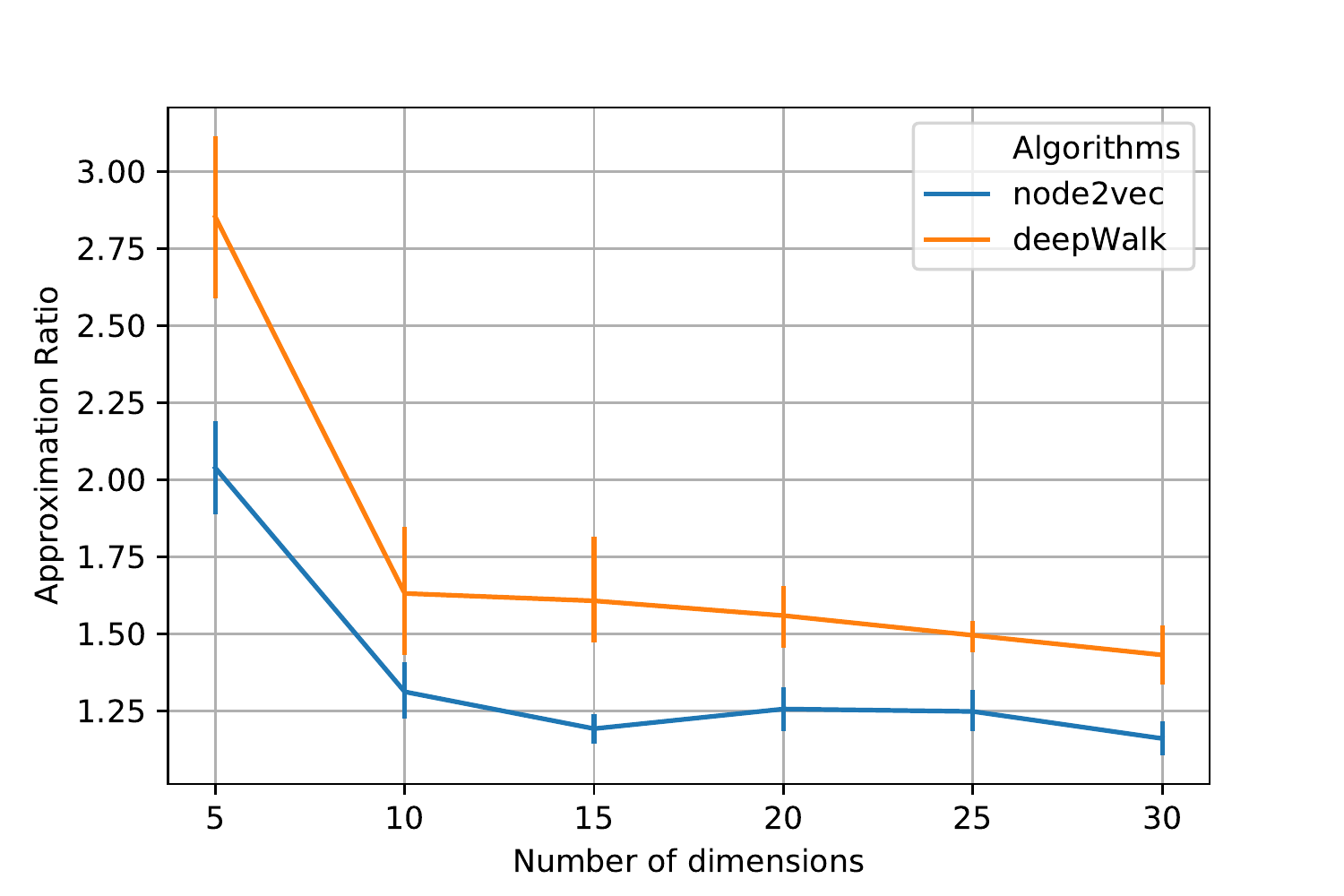}
\subcaption{}
\end{minipage}
\vspace{0.5cm}
\caption{Performance comparison of our proposed algorithm to that of greedy for (a) MCM (b) BM under adversarial model and (c) MCM under lomax model. (d)-(f) Parameter sensitivity of the proposed algorithm under adversarial model.}
\vspace{-0.5cm}
\label{fig:results}
\end{figure*} 
For synthetic datasets, we vary the distributions of edge weights. We compare the performance of our proposed heuristic algorithm to that of the greedy heuristic under two models based on generating edge weights. First, we consider an adversarial model. Under this model, edge weights are generated according to \cite{Reingold81} and as shown in Figure \ref{greedy-adversarial}. We set $n = 2^{t-1}$ with $t$ being the number of clusters of nodes and the largest distance between adjacent nodes is set to $3^{t-2}.$ Under the adversarial model, the greedy heuristic finds a solution with cost $O(n^{\log 3/2})$ times the optimum cost \cite{Reingold81}. Next, we consider the Lomax distribution (a long tail distribution) \cite{Lomax54} for generating the weights since they are widely used in real-world applications \cite{Wang2019}. We chose the Lomax distribution parameter from the set $\{2, 3, 5, 10, 50\}$. The parameter settings are motivated by \cite{Wang2019, Grover2016}. We set the number of walks per node and length of each random walk to $20$ for both deepWalk and node2vec. The embedding number of dimensions, return ($p$) and in-out parameter ($q$) are set to $10, 0.5$ and $2.0$ respectively. The experiments are repeated $5$ times for each parameter setting and the $95\%$ confidence interval plots are reported. If not mentioned, in our experiments, n = 100 and m = 4950 for non-bipartite graphs. We focus on MCM and BM algorithms in non-bipartite setting. We get similar results for bipartite graphs and hence omit them here. 


\subsection{Effect of Edge Weight Distribution}
We first present the results for MCM and BM under adversarial model as shown in Figures \ref{fig:results}(a) and (b) respectively. We plot the approximation ratios of various matching algorithms as a function of number of nodes in the graph. For both MCM and BM, the proposed node2vec and deepWalk based algorithms yield lower approximation ratio, hence better performance compared to greedy heuristic. Also, as the number of nodes increases, the performance of greedy heuristic decreases while that of embedding based techniques increases. One possible explanation for better performance of the embedding based techniques is due to their ability to learn the inherent one-dimensional structure of the adversarial model. However, the nature of the plots are reversed for MCM under the Lomax distribution model as shown Figure \ref{fig:results} (c). In this case, greedy beats both deepWalk and node2vec based algorithms. Note that, as the Lomax distribution parameter increases, both embedding based techniques gradually perform better and finally converge to the greedy solution.
\subsection{Parameter Sensitivity}
The embedding based matching algorithms involve a number of parameters. Below, we examine the effect of different parameters on the overall performance of embedding based techniques under adversarial model as shown in Figures \ref{fig:results} (d)-(f). All other parameters assume default values, except for the parameter being examined. We measure the approximation ratio as a function of number and length of walks per node, number of dimensions. The performance of both node2vec and deepWalk increase with the number, length of walks per node and number of dimensions as expected. Again, node2vec beats deepWalk across all values of parameters.

\section{Conclusion}\label{sec:conc}
In this paper, we developed approximate solutions for various matching problems on dense graphs. More precisely, we proposed a network embedding based heuristic algorithm using existing network embedding techniques. We also performed  simulations on synthetic datasets to obtain comparison results for empirical approximation ratios across different proposed and existing matching algorithms. Future directions include to consider the effect of other non-random walk based embedding techniques on  the overall performance of the proposed algorithm.
\section{Acknowledgment} 
This research was sponsored by the U.S. ARL and the U.K. MoD under Agreement Number W911NF-16-3-0001 and by the NSF under Grant CNS-1617437. The views and conclusions contained in this document are those of the authors and should not be interpreted as representing the official policies, either expressed or implied, of the National Science Foundation, U.S. ARL or the U.K. MoD. This document does not contain technology or technical data controlled under either the U.S. International Traffic in Arms Regulations or the U.S. Export Administration Regulations.
%
\bibliographystyle{abbrv}
\bibliography{refs} 


\end{document}